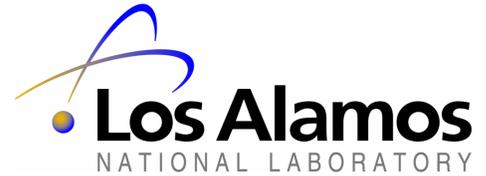

# Resolving Two Beams in Beam Splitters with a Beam Position Monitor

Sergey Kurennoy
LANSCE-1, Los Alamos National Laboratory
Los Alamos, NM 87545

*Abstract*

The beam transport system for the Advanced Hydrotest Facility (AHF) anticipates multiple beam splitters [1]. Monitoring two separated beams in a common beam pipe in the splitter sections imposes certain requirements on diagnostics for these sections. In this note we explore a two-beam system in a generic beam monitor and study the feasibility of resolving the positions of the two beams with a single diagnostic device.

# 1. Introduction.

In the Advanced Hydrotest Facility (AHF), 20-ns beam pulses (bunches) are extracted from the 50-GeV main proton synchrotron and then are transported to the target by an elaborated transport system [1]. The beam transport system splits the beam bunches into equal parts in its splitting sections so that up to 12 synchronous beam pulses can be delivered to the target for the multi-axis proton radiography. Information about the transverse positions of the beams in the splitters, and possibly the bunch longitudinal profile, should be delivered by some diagnostic devices. Possible candidates are the circular wall current monitors in the circular pipes connecting the splitter elements, or the conventional stripline BPMs [2]. In any case, we need some estimates on how well the transverse positions of the two beams can be resolved by these monitors.

To this end, let us consider a problem illustrated in Fig. 1. We make the following assumptions: (i) the vacuum chamber has an arbitrary cross section $S$ that does not change as the beams move along the chamber axis $z$; (ii) the chamber walls are perfectly conducting; and (iii) $(\omega b / \beta \gamma c)^2 \ll 1$, where $\omega$ is the frequency of interest, $b$ is a typical transverse dimension of the vacuum chamber, $\beta c$ is the beam velocity, and $\gamma = 1/\sqrt{1-\beta^2}$. The first condition means that the chamber cross section is the same in the vicinity of the diagnostic device, at least along a chamber segment a few times longer than $2b$. The last condition includes both the ultra relativistic limit, $\gamma \gg 1$, and the long-wavelength limit when, for a fixed $\gamma$, the wavelength of interest $\lambda \gg 2\pi b/\gamma$. For the AHF beam transport system with 50-GeV protons and $b$ on the order of 10 cm, the condition (iii) above is satisfied up to rather high frequencies, as high as $f = \omega/2\pi = 10$ GHz. As a result, the problem of calculating the beam transverse fields and the corresponding induced currents in the chamber walls is reduced to a two-dimensional electrostatic problem. Essentially the same problem was studied in Ref. [3], where an arbitrary transverse beam current distribution was considered. In our case, the beam charge distribution is just a sum of the two distributions, one with its center at the first beam position $\vec{r}_1$, and the other at $\vec{r}_2$. Using the same approach as [3], we calculate the transverse electric field created by the two beams at an arbitrary point $\vec{b}$ on the chamber wall.

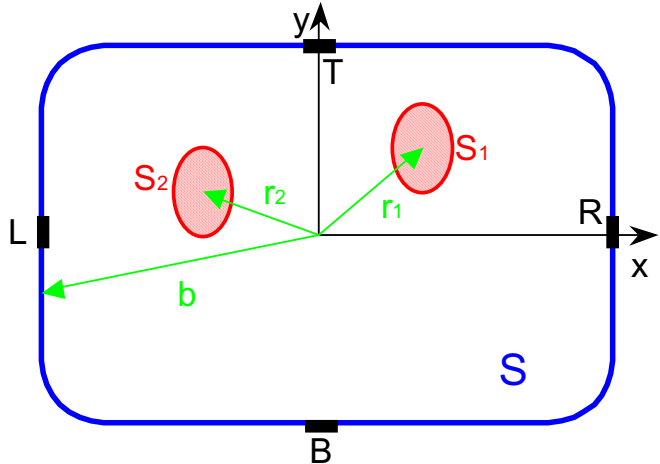

Figure 1. Transverse cross section of the vacuum chamber $S$ and the beams $S_1$ and $S_2$.

To simplify all calculations below, we first assume that both beams have the same (unit) charge per unit length and their charge distributions are axially symmetric around their axes. It was demonstrated that in the case of an axisymmetric charge distribution the field produced by the beam on the wall is the same as that of the pencil beam traveling along the distribution axis [3]. The beam-size corrections due to asymmetries of the beam current distributions are known [3], so that they can be included later. Also, to further simplify our consideration, let us consider the case of a circular cylindrical pipe.

## 2. Transverse Fields of Two Beams with Equal Currents.

In a circular cylindrical pipe of radius $b$, a pencil beam with the unit charge per unit length and a transverse offset $\vec{r}$ from the axis produces the following field at point $\vec{b}$ on the wall

$$e(\vec{r},\vec{b}) = \frac{1}{2\pi b} \frac{b^2 - r^2}{b^2 - 2\vec{b}\vec{r} + r^2}. \tag{2.1}$$

Due to the linearity, the field created at point $\vec{b}$ by two pencil beams, each with the unit charge per unit length, located at $\vec{r}_1$ and $\vec{r}_2$ is

$$E(\vec{r}_1, \vec{r}_2, \vec{b}) = e(\vec{r}_1, \vec{b}) + e(\vec{r}_2, \vec{b}). \tag{2.2}$$

Let us introduce two new vectors instead of $\vec{r}_1$ and $\vec{r}_2$: the center of the two-beam charge distribution $\vec{a} = (\vec{r}_1 + \vec{r}_2)/2$, and the beam spacing $\vec{d} = \vec{r}_1 - \vec{r}_2$. In terms of these vectors, the field (2.2) is

$$E(\vec{a},\vec{d};\vec{b}) = \frac{1}{2\pi b} \left[ \frac{b^2 - r_1^2}{b^2 - \vec{b}(2\vec{a}+\vec{d}) + r_1^2} + \frac{b^2 - r_2^2}{b^2 - \vec{b}(2\vec{a}-\vec{d}) + r_2^2} \right], \tag{2.3}$$

where $r_1^2 = a^2 + d^2/4 + \vec{a}\vec{d}$ and $r_2^2 = a^2 + d^2/4 - \vec{a}\vec{d}$. Since the four parameters ($a_x, a_y, d_x, d_y$) define the beam transverse positions completely, one needs, in principle, only to know the field (2.3) in four different points to find the beam positions. However, because the relations between the fields and the beam parameters are nonlinear, there is no guarantee that these four equations will allow us to find all 4 beam parameters from only 4 field measurements. Let us choose points **R**=(b,0), **T**=(0,b), **L**=(-b,0), **B**=(0,-b) as the points where the fields (or induced currents) are measured, see Fig. 1. These points can be locations of narrow striplines in a BPM or of resistors in a wall current monitor. Denoting $R = E(\vec{a},\vec{d};b,0)$, $T = E(\vec{a},\vec{d};0,b)$, and so on, and choosing $b=1$ in all equations below for simplicity (so that all transverse distances are now measured in units of $b$), we get

$$R = \frac{1}{2\pi} \left[ \frac{1 - r_1^2}{1 - 2a_x - d_x + r_1^2} + \frac{1 - r_2^2}{1 - 2a_x + d_x + r_2^2} \right], \tag{2.4}$$

$$T = \frac{1}{2\pi} \left[ \frac{1 - r_1^2}{1 - 2a_y - d_y + r_1^2} + \frac{1 - r_2^2}{1 - 2a_y + d_y + r_2^2} \right], \tag{2.5}$$

$$L = \frac{1}{2\pi} \left[ \frac{1 - r_1^2}{1 + 2a_x + d_x + r_1^2} + \frac{1 - r_2^2}{1 + 2a_x - d_x + r_2^2} \right], \tag{2.6}$$

$$B = \frac{1}{2\pi} \left[ \frac{1 - r_1^2}{1 + 2a_y + d_y + r_1^2} + \frac{1 - r_2^2}{1 + 2a_y - d_y + r_2^2} \right]. \tag{2.7}$$

One can try to solve these simultaneous equations numerically with respect to $a_x, a_y, d_x, d_y$ for given $R, T, L, B$, e.g. by minimization methods, but we should expect to get more insight by studying their approximate analytical solutions.

Let us first consider <u>a particular case</u> of $\vec{a} = 0$, i.e. the center of the two-beam charge distribution is on the chamber axis. Then $r_1^2 = r_2^2 = d^2/4$, and Eqs. (2.4-7) are reduced to

$$R = L = \frac{1}{\pi} \frac{1-(d/2)^4}{1+(d/2)^4 - (d_x^2 - d_y^2)/2}, \tag{2.8}$$

$$T = B = \frac{1}{\pi} \frac{1-(d/2)^4}{1+(d/2)^4 + (d_x^2 - d_y^2)/2}. \tag{2.9}$$

Combining these equations, one can get

$$\frac{R-T}{R+T} = \frac{1}{2} \frac{d_x^2 - d_y^2}{1+(d/2)^4}. \tag{2.10}$$

One should note that up to this point there were no assumptions made on the magnitude of $d_x, d_y$, and Eqs. (2.8-9), as well as Eq. (2.10), are exact for $\vec{a} = 0$. In the AHF beam splitters the beams are well separated in the horizontal plane, and one should expect that $d_x^2 \gg d_y^2$, as well as $d_x^2 \gg a_x^2, d_x^2 \gg a_y^2$, but definitely $|d_x| < 2$ (in units of $b$). It is reasonable to consider $|d_x| \leq 1$, so that $(d/2)^4 \leq 1/16 \ll 1$, and the ratio (2.10) is then approximately proportional to the beam separation squared, $(R-T)/(R+T) \cong d_x^2/2$.

Another <u>particular case</u> when the final results can be explicitly derived from Eqs. (2.4-7), is the case of $a, d \ll 1$. Performing Taylor expansions in Eqs. (2.4-7), one obtains

$$\binom{R}{L} = \frac{1}{2\pi} \left[ 2 \pm 4a_x + 4(a_x^2 - a_y^2) + (d_x^2 - d_y^2) \pm a_x (4a_x^2 - 12a_y^2 + 3d_x^2 - 3d_y^2) \mp 6a_y d_x d_y + O_h(\delta^4) \right], \tag{2.11}$$

$$\binom{T}{B} = \frac{1}{2\pi} \left[ 2 \pm 4a_y - 4(a_x^2 - a_y^2) - (d_x^2 - d_y^2) \mp a_y (12a_x^2 - 4a_y^2 + 3d_x^2 - 3d_y^2) \mp 6a_x d_x d_y + O_v(\delta^4) \right], \tag{2.12}$$

where the top (bottom) signs correspond to the top (bottom) variable in the LHS. Combining these equations, we have

$$R - L = \frac{1}{\pi} \left[ 4a_x + a_x (4a_x^2 - 12a_y^2 + 3d_x^2 - 3d_y^2) - 6a_y d_x d_y + O(\delta^5) \right], \tag{2.13}$$

$$T - B = \frac{1}{\pi} \left[ 4a_y - a_y (12a_x^2 - 4a_y^2 + 3d_x^2 - 3d_y^2) - 6a_x d_x d_y + O(\delta^5) \right], \tag{2.14}$$

$$(R+L) - (T+B) = \frac{2}{\pi} \left[ 4(a_x^2 - a_y^2) + (d_x^2 - d_y^2) + O(\delta^4) \right], \tag{2.15}$$

$$S \equiv R + L + T + B = \frac{1}{\pi} \left[ 4 + O(\delta^4) \right]. \tag{2.16}$$

One can see that the following field (signal) ratios are convenient for characterizing the two-beam system:

$$\frac{R-L}{S} = a_x + a_x \left[ a_x^2 - 3a_y^2 + \frac{3}{4}(d_x^2 - d_y^2) \right] - \frac{3}{2} a_y d_x d_y + O(\delta^5), \tag{2.17}$$

$$\frac{T-B}{S} = a_y - a_y \left[ 3a_x^2 - a_y^2 + \frac{3}{4}(d_x^2 - d_y^2) \right] - \frac{3}{2} a_x d_x d_y + O(\delta^5), \tag{2.18}$$

$$\frac{(R+L)-(T+B)}{S} = 2(a_x^2 - a_y^2) + \frac{1}{2}(d_x^2 - d_y^2) + O(\delta^4). \tag{2.19}$$

Obviously, the first two ratios, Eqs. (2.17-18), give information about the position of the center of the combined charge distribution for two beams, while the last one, Eq. (2.19), can be used to extract info on the beam separation. If we assume the beams are separated mainly in the horizontal plane, i.e. $d_x^2 \gg d_y^2$, $d_x^2 \gg a_x^2$, $d_x^2 \gg a_y^2$, then this ratio behaves similarly to the one in Eq. (2.10) above: $[(R+L)-(T+B)]/S \cong d_x^2/2$. One should remind, however, that results (2.11-19) are valid only for the case of $d_x^2 \ll 1$.

The results above are derived under the assumption that the two beams are pencil-like, or equivalently, that they both have an axisymmetric beam charge distributions. Let us now consider how these results change if the transverse distributions of the beam current are not axisymmetric, i.e. let us find the beam-size corrections. For example, assume first that both beams have the same normalized double-Gaussian charge distribution in their transverse cross sections,

$$\lambda(x, y) = \frac{1}{2\pi\sigma_x\sigma_y} \exp\left[-\frac{(x-x_0)^2}{2\sigma_x^2} - \frac{(y-y_0)^2}{2\sigma_y^2}\right], \tag{2.20}$$

where $(x_0, y_0)$ are the beam-center coordinates, and $\sigma_x, \sigma_y$ define its transverse size. Of course, it is natural to consider $\sigma_x, \sigma_y \ll 1$ (in units of the chamber radius $b$). From Eq. (5) of Ref. [3] one can easily derive Eqs. (2.11-12) with the beam-size corrections. The corrections to Eqs. (2.11-12) are as follows:

$$\begin{pmatrix} \Delta R \\ \Delta L \end{pmatrix} = \frac{1}{2\pi}\left[4M_2 \pm 12a_x M_2 + O(\delta^4)\right], \tag{2.21}$$

$$\begin{pmatrix} \Delta T \\ \Delta B \end{pmatrix} = \frac{1}{2\pi}\left[-4M_2 \mp 12a_y M_2 + O(\delta^4)\right], \tag{2.22}$$

where the second-order moment $M_2 = \sigma_x^2 - \sigma_y^2$ for the double-Gaussian distribution (2.20).
One can see that the beam-size corrections will modify Eqs. (2.13-15) into

$$R - L = \frac{1}{\pi}\left\{4a_x + a_x\left[4a_x^2 - 12a_y^2 + 3(d_x^2 - d_y^2) + 12M_2\right] - 6a_y d_x d_y + O(\delta^5)\right\}, \tag{2.23}$$

$$T - B = \frac{1}{\pi}\left\{4a_y - a_y\left[12a_x^2 - 4a_y^2 + 3(d_x^2 - d_y^2) + 12M_2\right] - 6a_x d_x d_y + O(\delta^5)\right\}, \tag{2.24}$$

$$(R+L)-(T+B) = \frac{2}{\pi}\left[4(a_x^2 - a_y^2) + (d_x^2 - d_y^2) + 4M_2 + O(\delta^4)\right], \tag{2.25}$$

while Eq. (2.16) remains unchanged except for the higher order corrections. As a result, the corrected signal ratios (2.17-19) become

$$\frac{R-L}{S} = a_x + a_x\left[a_x^2 - 3a_y^2 + \frac{3}{4}(d_x^2 - d_y^2) + 3M_2\right] - \frac{3}{2}a_y d_x d_y + O(\delta^5), \tag{2.26}$$

$$\frac{T-B}{S} = a_y - a_y\left[3a_x^2 - a_y^2 + \frac{3}{4}(d_x^2 - d_y^2) + 3M_2\right] - \frac{3}{2}a_x d_x d_y + O(\delta^5), \tag{2.27}$$

and

$$\frac{(R+L)-(T+B)}{S} = 2(a_x^2 - a_y^2) + \frac{1}{2}(d_x^2 - d_y^2) + 2M_2 + O(\delta^4). \tag{2.28}$$

It should be noted at this point that the results above can be generalized to more general transverse beam charge distributions. As was demonstrated in [3], within the framework of our consideration, all beam-size corrections (shown in blue in Eqs. (2.23-28)) enter the expressions for fields and signals via their multipole moments. In particular, all corrections up to the third order include only the quadrupole moment $M_2$ of the charge distribution when the charge distribution has two axis of symmetry [3]. So, the results (2.21-28) above can be applied for other symmetric beam charge distributions with the corresponding $M_2$ substituted. For example, if one considers a uniform beam with a rectangular cross section $2\sigma_x \times 2\sigma_y$:

$$\lambda(x, y) = \theta(x - x_0 + \sigma_x)\theta(\sigma_x - x + x_0)\theta(y - y_0 + \sigma_y)\theta(\sigma_y - y + y_0)/(4\sigma_x\sigma_y), \tag{2.29}$$

where $\theta(x)$ is the step function, and $(x_0, y_0)$ are the beam-center coordinates, the value of its quadrupole moment $M_2 = (\sigma_x^2 - \sigma_y^2)/3$ should be used. In fact, according to beam simulations [4], the transverse cross section of the proton beams after splitting (beamlets) in the AHF beam splitters is close to a semi-ellipse with a double-Gaussian charge distribution, as illustrated in Fig. 2.

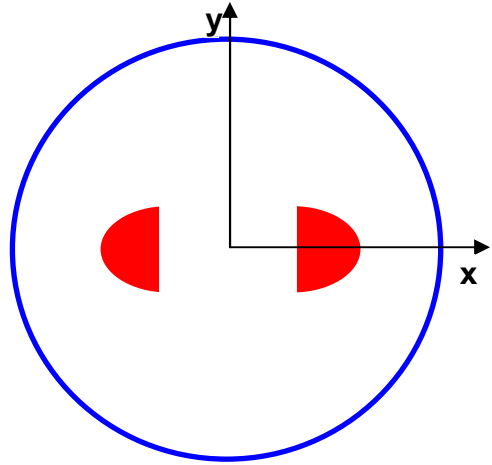

Figure 2. Transverse cross section of the vacuum chamber (blue) and of two split beams (red) in AHF beam splitters.

Relevant information on the moments of a few transverse beam-charge distributions is summarized below in Table 1. The last column shows, for non-symmetric distributions, the distance of the beam center from the cut edge. The beam center is assumed to be at $(x_0, y_0)$.

Table 1. Definition and some properties of some transverse beam-charge distributions

| Distribution | $\lambda(x, y)$ | $M_2$ | $\Delta_x$ |
|---|---|---|---|
| Double Gaussian | $\dfrac{1}{2\pi\sigma_x\sigma_y}\exp\left[-\dfrac{(x-x_0)^2}{2\sigma_x^2} - \dfrac{(y-y_0)^2}{2\sigma_y^2}\right]$ | $\sigma_x^2 - \sigma_y^2$ | n/a |
| Uniform rectangular | $\theta(x-x_0+\sigma_x)\theta(\sigma_x-x+x_0) \times$ $\theta(y-y_0+\sigma_y)\theta(\sigma_y-y+y_0)/(4\sigma_x\sigma_y)$ | $\dfrac{1}{3}(\sigma_x^2 - \sigma_y^2)$ | n/a |
| Uniform rhs semicircle | $\dfrac{2}{\pi\sigma^2}\theta\left[\sigma^2 - (x-x_0)^2 - (y-y_0)^2\right]\theta(x-x_0+\Delta_x)$ | $-\dfrac{16}{9\pi^2}\sigma^2$ | $\dfrac{4}{3\pi}\sigma$ |
| Semi-elliptic rhs double Gaussian | $\dfrac{1}{\pi\sigma_x\sigma_y}\exp\left[-\dfrac{(x-x_0+\Delta_x)^2}{2\sigma_x^2} - \dfrac{(y-y_0)^2}{2\sigma_y^2}\right]\theta(x-x_0+\Delta_x)$ | $\sigma_x^2 - \sigma_y^2 - \dfrac{2\sigma_x^2}{\pi}$ | $\sqrt{\dfrac{2}{\pi}}\sigma_x$ |

One should note that in a general case of two non-symmetric beam-charge distributions, there will be corrections of the order of $O(\delta^3)$ in Eq. (2.21), and the resulting corrections will enter Eqs. (2.23,25), and (2.26,28). However, for two half-beams shown in Fig. 2, the corrections due to the next order moment $M_3$ cancel each other if the charge in each of the beamlets is distributed symmetrically with respect to the horizontal plane, and if their currents are the same.

If we consider now our case of interest, namely $|a_x|,|a_y|,|d_y|,|\sigma_x|,|\sigma_y| \ll |d_x| \ll 1$, where the last inequality reflects the assumptions used in deriving Eqs. (2.23-28), the leading terms in the ratios (2.26-28) are

$$(R-L)/S \cong a_x(1+3d_x^2/4); \quad (T-B)/S \cong a_y(1-3d_x^2/4); \quad [(R+L)-(T+B)]/S \cong d_x^2/2. \qquad (2.30)$$

The last ratio allows us to find the separation of the two beams in the horizontal plane, while the first two provide information on deviations of the two-beam charge distribution center from the chamber axis. The corrections to these leading terms are shown explicitly in Eqs. (2.26-28), and are expected to be small. Next we move to lift the only unpractical limitation left, $|d_x| \ll 1$, in the results above.

In the case when all possible beam separations, $|d_x| < 2$, are considered, expanding the field equations (2.4-7) in terms of small parameters $a, d_y$ becomes more involved. To simplify notations, we denote $\vec{h} = \vec{d}/2$, so that $-1 < h_x < 1$. Shown below are the two lowest terms in the signal ratio series (checked with the symbolic algebra package in *Mathematica* [5]):

$$\frac{R-L}{S} = a_x \frac{(1+h_x^2)^2}{(1-h_x^2)(1+h_x^4)} - 2a_y h_y \frac{h_x(1+h_x^2)(3+h_x^2)}{(1-h_x^2)^2(1+h_x^4)} + O(\delta^3), \qquad (2.31)$$

$$\frac{T-B}{S} = a_y \frac{(1-h_x^2)^2}{(1+h_x^2)(1+h_x^4)} - 2a_x h_y \frac{h_x(1-h_x^2)(3-h_x^2)}{(1+h_x^2)^2(1+h_x^4)} + O(\delta^3), \qquad (2.32)$$

$$\frac{(R+L)-(T+B)}{S} = 2\frac{h_x^2}{1+h_x^4} + 2\frac{1+3h_x^2}{(1+h_x^4)^2}(a_x^2 - a_y^2 - h_y^2) + O(\delta^3). \qquad (2.33)$$

One can see that for small $h_x$ these equations reproduce the corresponding parts of expansions (2.26-28). The beam-size contributions here are exactly the same as in Eqs. (2.26-28).

## 3. Transverse Fields of Two Beams with Unequal Beam Currents.

Let us now assume that while the total beam current is fixed, the currents in two split beams are not equal. Then instead of Eq. (2.2), the transverse field at point $\vec{b}$ on the wall due to two unequal beams located at $\vec{r}_1$ and $\vec{r}_2$ will be

$$E(\vec{r}_1, \vec{r}_2, \vec{b}) = \lambda_1 e(\vec{r}_1, \vec{b}) + \lambda_2 e(\vec{r}_2, \vec{b}), \qquad (3.1)$$

where the weights $\lambda_1, \lambda_2$ are constrained by $\lambda_1 + \lambda_2 = 2$. Obviously, for two identical beams considered in Sect. 2, $\lambda_1 = \lambda_2 = 1$. It is convenient to introduce one parameter of charge misbalance $k$ by choosing $\lambda_1 = 2k, \lambda_2 = 2(1-k)$, where $0 < k < 1$, with $k = 1/2$ corresponding to two equal beams. Now the center of the two-beam charge distribution $\vec{a} = k\vec{r}_1 + (1-k)\vec{r}_2$, and with the beam spacing $\vec{d} = \vec{r}_1 - \vec{r}_2$, we have $\vec{r}_1 = \vec{a} + (1-k)\vec{d}$, and $\vec{r}_2 = \vec{a} - k\vec{d}$. Then in terms of vectors $\vec{a}, \vec{d}$, the field (3.1) is

$$E(\vec{a},\vec{d};\vec{b}) = \frac{1}{\pi b}\left[ k\frac{b^2 - r_1^2}{b^2 - 2\vec{b}\left(\vec{a}+(1-k)\vec{d}\right)+r_1^2} + (1-k)\frac{b^2 - r_1^2}{b^2 - 2\vec{b}\left(\vec{a}-k\vec{d}\right)+r_1^2} \right], \tag{3.2}$$

where $r_1^2 = a^2 + (1-k)^2 d^2 + 2(1-k)\vec{a}\vec{d}$ and $r_2^2 = a^2 + k^2 d^2 - 2k\vec{a}\vec{d}$, cf. Eq. (2.3). Equations (2.4-7) for the fields $R, T, L, B$ in the chosen points (with all transverse dimensions expressed in units of $b$) are modified accordingly.

When the beam offsets are small compared to the chamber radius, i.e. $|a_x|,|a_y|,|d_x|,|d_y| \ll 1$ (again using $b = 1$), we obtain instead of Eqs. (2.17-19)

$$\frac{R-L}{S} = a_x + a_x\left[a_x^2 - 3a_y^2 + 3k(1-k)(d_x^2 - d_y^2)\right] - 6k(1-k)a_y d_x d_y$$
$$+ k(1-k)(1-2k)d_x(d_x^2 - 3d_y^2) + O(\delta^4), \tag{3.3}$$

$$\frac{T-B}{S} = a_y + a_y\left[a_y^2 - 3a_x^2 - 3k(1-k)(d_x^2 - d_y^2)\right] - 6k(1-k)a_x d_x d_y$$
$$+ k(1-k)(1-2k)d_y(d_y^2 - 3d_x^2) + O(\delta^4), \tag{3.4}$$

and

$$\frac{(R+L)-(T+B)}{S} = 2(a_x^2 - a_y^2) + 2k(1-k)(d_x^2 - d_y^2) + O(\delta^4). \tag{3.5}$$

As for the beam-spot corrections, they can be obtained from results [3] in a way similar to that used above. The beam-size corrections to Eqs. (3.3-5) for the case of two different arbitrary symmetric distributions of the beam current in the beam transverse cross sections, can be obtained by replacing the size correction terms (highlighted in blue) in Eqs. (2.23-28) as follows:

$$M_2 \to kM_2^{(1)} + (1-k)M_2^{(2)}, \tag{3.6}$$

i.e., the combination $kM_2^{(1)} + (1-k)M_2^{(2)}$ of the quadrupole moments of two beams should substitute the quadrupole moment $M_2$. Obviously, when $k \to 1/2$, i.e. the beams are identical, the equations (3.3-5) are reduced to Eqs. (2.17-19), or to Eqs. (2.26-28), if the beam-size corrections are taken into account.

For the case of interest, $|a_x|,|a_y|,|d_y| \ll |d_x| \ll 1$, and when $k \neq 1/2$, the leading terms in the above ratios are

$$(R-L)/S \cong k(1-k)(1-2k)d_x^3 + a_x\left[1 + 3k(1-k)d_x^2\right];$$
$$(T-B)/S \cong a_y\left[1 - 3k(1-k)d_x^2\right]; \text{ and } \left[(R+L)-(T+B)\right]/S \cong 2k(1-k)d_x^2. \tag{3.7}$$

These expressions should be compared with Eqs. (2.30). The last two ratios will be close to those from Eqs. (2.30) as long as the charge misbalance is not too big, while the first one receives a correction (the first term) that can make extracting the horizontal position of the two-beam charge center from the signal ratio measurements inaccurate.

For an arbitrary beam separation, $|d_x|<2$, and with $a,|d_y|\ll 1$ (again we use $\vec{h}=\vec{d}/2$, so that $-1<h_x<1$), perturbation expansions of the signal ratios become rather cumbersome. Below we show the leading term in one of the ratios obtained with the *Mathematica* symbolic algebra package:

$$\frac{(R+L)-(T+B)}{S}=8k(1-k)h_x^2\frac{1-16k(1-k)\left[1-3k(1-k)\right]h_x^4}{1-16(1-2k)^2\left[1-2k(1-k)\right]h_x^4-256k^4(1-k)^4h_x^8}+O(\delta). \quad (3.8)$$

The expressions for the other two ratios are too complicated to be useful. Let us now introduce another parameter for the charge misbalance $\varepsilon=k-1/2$ ($-1/2<\varepsilon<1/2$, with $\varepsilon=0$ corresponding to two identical beams), and give the lowest terms of the ratio expansions for arbitrary $|h_x|<1$, and $a,|h_y|,|\varepsilon|\ll 1$:

$$\frac{R-L}{S}=a_x\frac{(1+h_x^2)^2}{(1-h_x^2)(1+h_x^4)}-4\varepsilon h_x^2\frac{1+h_x^2}{(1-h_x^2)(1+h_x^4)}+O(\delta^2), \quad (3.9)$$

$$\frac{T-B}{S}=a_y\frac{(1-h_x^2)^2}{(1+h_x^2)(1+h_x^4)}+O(\delta^2), \quad (3.10)$$

$$\frac{(R+L)-(T+B)}{S}=2\frac{h_x^2}{1+h_x^4}+\frac{2}{(1+h_x^4)^2}\left[\begin{array}{c}(1+3h_x^2)(a_x^2-a_y^2-h_y^2)-\\ -16\varepsilon a_x h_x^5-4\varepsilon^2 h_x^2(1-5h_x^4)\end{array}\right]+O(\delta^3). \quad (3.11)$$

These results are to be compared with those given by Eqs. (2.31-33). One should note that ratio (3.9) gets a correction already in the leading order, while for (3.10) and especially (3.11) one can expect noticeable corrections only for large values of the beam separation and / or of charge misbalance.

## 4. Results for Relevant Values of Parameters.

According to the preliminary design of the AHF splitter sections [1,4], the two split beamlets look as shown in Fig. 2. The horizontal beam separation $g$, from one cut edge to the other, increases from 5 mm near the entrance of the pulsed magnetic septum to about 5.2 cm at the entrance of the first DC magnetic septum. From beam simulations [4], the transverse beam-charge distribution in the beamlets can be approximated as the semi-elliptic double-Gaussian one, cf. Tab. 1, with rms values $\sigma_x=3.7$ mm, and $\sigma_y=2.4$ mm. The vacuum pipe ID changes from 2" to 4". Since the center-to-center horizontal beam separation $d_x=g+2\Delta_x$, see in Tab. 1, the ratio $|d_x/b|$ varies from rather small values to above 1 along the beam splitter. For example, at the entrance of the pulsed magnetic septum $d_x\simeq 10.8$ mm and $b=25.4$ mm, so that $d_x/b\simeq 0.43$; near the first DC magnetic septum $d_x\simeq 57.9$ mm, $b=50.8$ mm, and $d_x/b\simeq 1.14$. We will assume $|a_x,a_y,d_y|\leq 0.2b$ and $|\sigma_x,\sigma_y|\leq 0.15b$ for our estimates. Two kinds of the beam splitters are anticipated in the AHF beam transport system: most will produce equal beamlet currents (1:1 splitters), but some may be needed to split the input beam with the current ratio 2:1. While the ratio of split beam intensities will vary, it is reasonable to expect it to be near 1 or 2, respectively, within a few percent. Based on these values, we explore the misbalance parameter $k$ in the range of 0.3-0.7. Some results for the signal ratios are presented in Figs. 3-6.

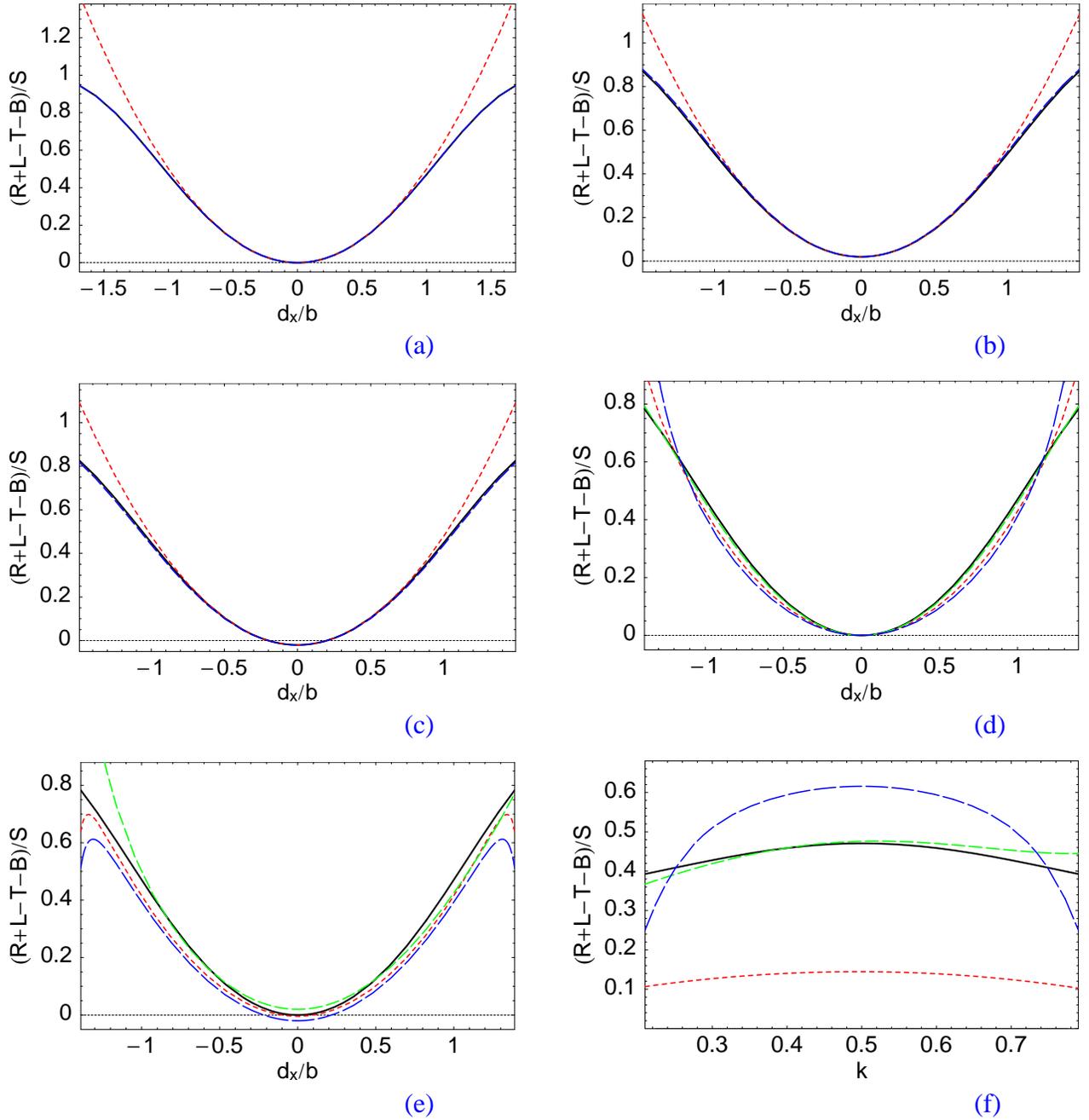

Figure 3. Signal ratio $(R + L - T - B)/S$: (a) for two identical beams versus $d_x/b$ with $a_x = 0, a_y = 0, d_y = 0$: exact, Eqs. (2.4-7) (black solid curve), Eq. (2.19) (red short-dashed), Eq. (2.33) (blue dashed); (b) same for $a_x = 0.1b, a_y = 0, d_y = 0$; (c) same for $a_x = 0, a_y = 0, d_y = 0.2b$; (d) exact, Eq. (3.2), for two unequal beams with $a_x = 0, a_y = 0, d_y = 0$: black solid curve – $k = 0.5$ (two equal beams), $k = 0.3$ – red short-dashed, $k = 0.6$ – green dashed, $k = 0.75$ – blue long-dashed; (e) same with $k = 0.5, a_x = 0, a_y = 0, d_y = 0$ (black solid), with $k = 0.3, a_x = 0, a_y = 0, d_y = 0.1b$ (red short-dashed), with $k = 0.7, a_x = 0.1b, a_y = 0, d_y = 0$ (green dashed), with $k = 0.7, a_x = 0, a_y = -0.1b, d_y = 0$ (blue long-dashed); (f) versus $k$ for $a_x = 0, a_y = 0, d_x = b, d_y = 0$ (black solid curve), for $a_x = 0.1b, a_y = 0, d_x = 0.5b, d_y = 0$ (red short-dashed), for $a_x = 0, a_y = -0.05b, d_x = b, d_y = 0$ (green dashed), for $a_x = 0, a_y = 0, d_x = 1.2b, d_y = 0.2b$ (blue long-dashed).

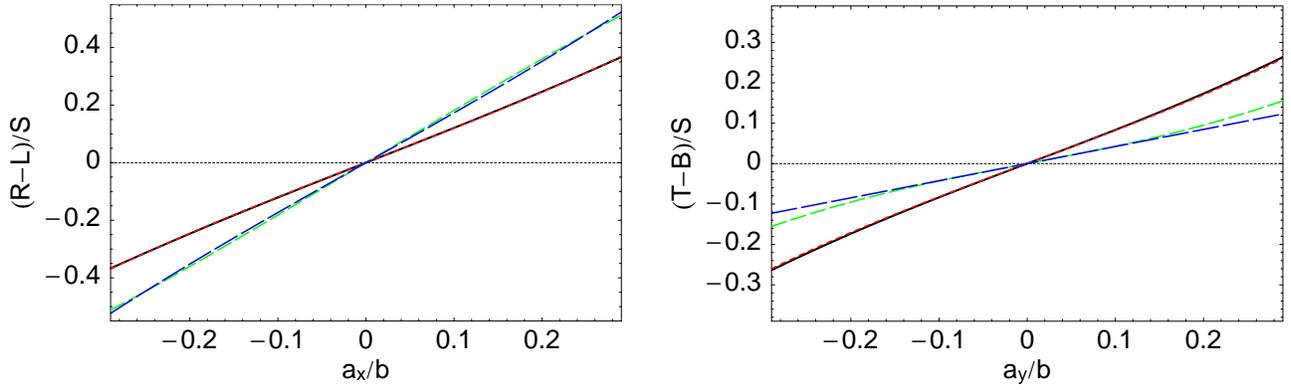

Figure 4. Signal ratios $(R-L)/S$ versus $a_x/b$ and $(T-B)/S$ versus $a_y/b$ for two equal beams: exact with $a_{y,x}=0, d_x=0.5b, d_y=0$ from Eqs. (2.4-7) (black solid curve) and with $a_{y,x}=0.1b, d_x=b, d_y=0$ (green dashed), compared to analytical approximations (2.17-18) (red short-dashed and blue dashed, correspondingly).

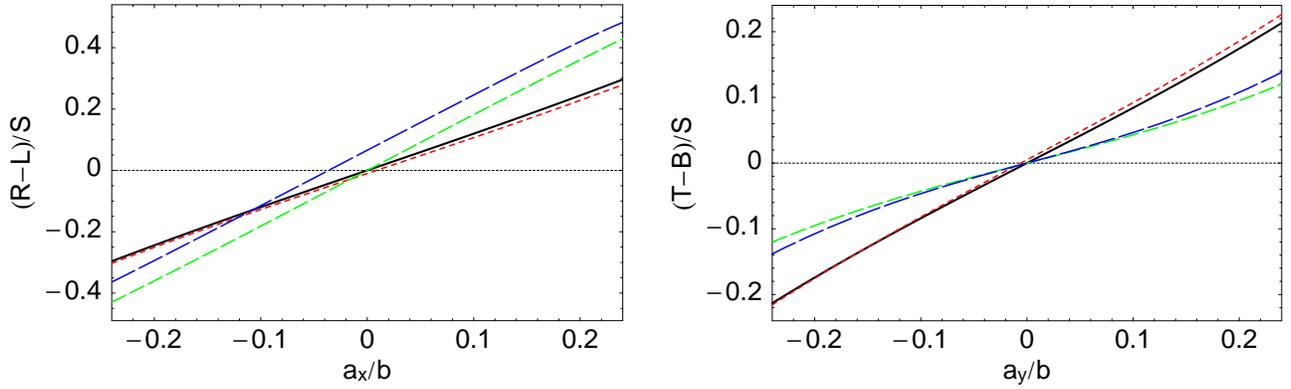

Figure 5. Signal ratios $(R-L)/S$ versus $a_x/b$ and $(T-B)/S$ versus $a_y/b$ for two beams, exact: with $a_{y,x}=0, d_x=0.5b, d_y=0.1b$ for $k=0.5$ (black solid curve) compared to $k=0.7$ (red short-dashed); and with $a_{y,x}=0.1b, d_x=b, d_y=0$ for $k=0.5$ (equal beams, green dashed) compared to $k=0.4$ (blue long-dashed).

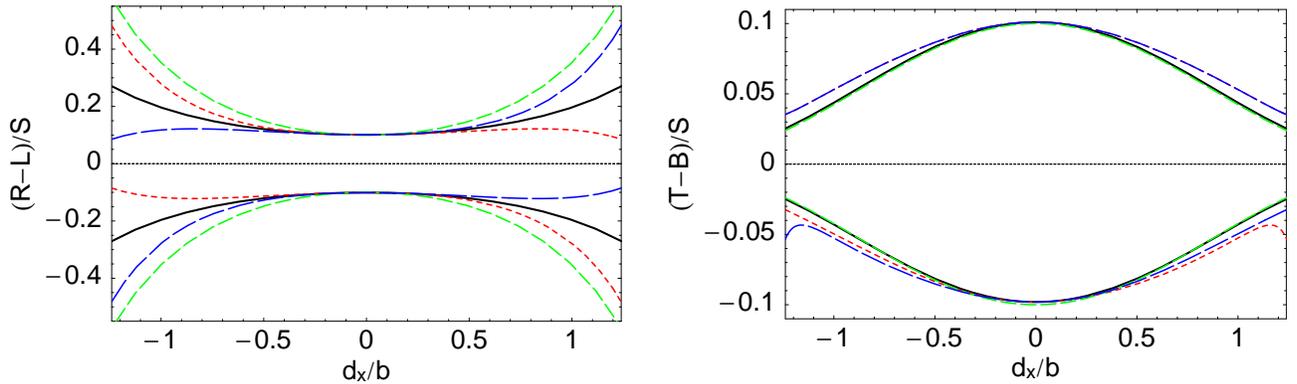

Figure 6. Signal ratios $(R-L)/S$ and $(T-B)/S$ versus $d_x/b$ for two beams, with fixed values of $a_x=0.1b$ (top group, left), $a_x=-0.1b$ (bottom, left), or $a_y=0.1b$ (top, right), $a_y=-0.1b$ (bottom, right): exact, $k=0.5$ (black solid); exact, $k=0.7$ (red short-dashed) compared to Eqs. (3.9-10) for $k=0.7$ (green dashed); and exact, $k=0.3$ (blue long-dashed).

As one can see from Figs. 3, the horizontal beam separation can be found reliably from the quadrupole signal ratio $(R+L-T-B)/S$ for reasonably large values of $|d_x/b|$ within the required range of beam charge misbalance. While we have not introduced beam-size effects in Figs. 3, these effects will not change the results significantly, according to Eqs. (2.28) and (3.6).

The vertical position of the beam charge center $a_y$ can also be determined from the measured signal ratio $(T-B)/S$, even for large values of the beam separation and of the current misbalance, cf. Figs. 4-6, especially when the measured value of the beam separation is taken into account. On the other hand, Figs. 5-6 show that it can be difficult to find accurately the horizontal position of the beam charge center $a_x$ when $|d_x/b|$ is large and the beam currents are far from being equal. However, if the current ratio is known from independent measurements, e.g., with current monitors, the value of $a_x$ can be restored.

For two points in the AHF splitters that were chosen above, the expected value of the quadrupole signal ratio $q \equiv (R+L-T-B)/S$ is: at the entrance of the pulsed magnetic septum, when $d_x/b \simeq 0.43$, $q = 0.088$ for $k = 1/2$ and $q = 0.078$ for $k = 1/3$; and near the first DC magnetic septum, when $d_x/b \simeq 1.14$, $q = 0.586$ for $k = 1/2$ and $q = 0.398$ for $k = 1/3$. It is worth mentioning that beam-size corrections (included in the $q$-values above) are small in both points: -0.003 and -0.0008, respectively.

One important conclusion from Figs. 3-6 is that approximate formulas for the signal ratios (2.17-19), (2.31-33), (3.3-5), and (3.9-11) are accurate enough in their regions of applicability. These formulas allow us to derive the parameters $a_x$, $a_y$, and $d_x$ of the two-beam system from post-processing the signal measurements with a simple four-stripline BPM in the AHF beam splitters within a few percent, when beam-size corrections are taken into account. One should note that all the derivations above assumed narrow striplines in BPMs, for simplicity. Should a finite azimuthal width $\phi$ of BPM electrodes be taken into account, one has to integrate the transverse beam fields along the electrode width. This would lead to usual additional form-factors like $\sin(\phi/2)/(\phi/2)$ in the signal ratios, see e.g. in Ref. [3]. When $\phi \ll \pi/2$, all these form-factors tend to 1.

## 5. Summary.

The transverse fields produced by two separated beams in a vacuum chamber have been calculated. It was demonstrated in the case of a relatively large horizontal beam separation in a circular vacuum chamber that combining signal (induced current) measurements in four points on the chamber walls allows us to reliably determine the horizontal beam separation $d_x$ from the ratio $(R+L-T-B)/S$, even if two beams have intensities that differ by a factor of 2, cf. Eqs. (3.11), (3.8), and (3.7). Finding the vertical position $a_y$ of the beam charge center is also relatively easy from the signal ratio $(T-B)/S$, see Eqs. (3.10) and (3.7). However, if no additional information on the current ratio of two beams (e.g., from beam current monitors somewhere in downstream sections where the two beams are in two different beam pipes) is available, it will be difficult to recover the horizontal position $a_x$ of the beam charge center from the signal ratios, see (3.9), (3.7). As for the vertical separation $d_y$ of the beams, it enters the leading terms of the ratios only in the combination $d_x^2 - d_y^2$ or in the higher order terms, so it will impossible to find its value from the signal ratios in the present pickup configuration if we expect $d_x^2 \gg d_y^2$. For the particular parameters of the AHF beam splitters, when the ratio of beam currents is known by design or from independent measurements with beam current monitors, one can derive the parameters $a_x$, $a_y$, and $d_x$ of the two-beam system from post-processing the signal measurements with a simple four-stripline BPM within an accuracy of a few percent. The beam-size corrections calculated for this case based on results of beam-dynamics simulations [4] are small. These conclusions can be checked using bench measurements with two wires [2].

In conclusion, one should emphasize that only a specific case of two well-separated beams in a simple pickup system is explored. A more general analysis with a multistrip monitor using discrete FFT signal processing (e.g., see [6]) can recover higher geometrical moments of the beam charge distribution. Such a system can be combined with a wall current monitor, but it will be definitely much more complicated.

## 6. References.